\documentclass[12pt]{iopart}

\expandafter\let\csname equation*\endcsname\relax
\expandafter\let\csname endequation*\endcsname\relax 

\usepackage[english]{babel}
\usepackage{amsmath}
\usepackage{amssymb}
\usepackage{graphicx}
\usepackage{color}

\begin{document}

\title{Diffusion in periodic, correlated random forcing landscapes}
\author{David S. Dean,$^1$ Shamik Gupta,$^2$ Gleb Oshanin,$^{3,4}$ Alberto
Rosso,$^2$ Gr\'egory Schehr$^2$}
\address{$^1$Universit\'e de Bordeaux and CNRS, Laboratoire 
d'Ondes et Mati\`{e}re d'Aquitaine (LOMA), UMR 5798, F-33400 Talence, France,\\
$^2$Laboratoire de Physique Th{\'e}orique et Mod{\`e}les  Statistiques
(UMR CNRS 8626), Universit\'e de Paris-Sud, Orsay Cedex, France, \\ $^3$Sorbonne Universit\'es, UPMC Univ Paris 06, UMR 7600, LPTMC, F-75005,
Paris, France, \\ $^4$CNRS, UMR 7600, Laboratoire de Physique
Th\'{e}orique de la Mati\`{e}re Condens\'{e}e, F-75005, Paris, France}
\date{\today}

\begin{abstract}
We study the dynamics of a Brownian particle in a strongly correlated quenched random  potential defined as a periodically-extended (with period $L$) finite trajectory of a
fractional Brownian motion with arbitrary Hurst exponent $H \in (0,1)$. 
While the periodicity ensures that the ultimate long-time
behavior is diffusive, the generalised 
Sinai potential considered here leads to a strong logarithmic confinement 
of 
particle trajectories at intermediate times.
These two competing trends lead to dynamical frustration and result in
a rich statistical behavior of the diffusion coefficient $D_L$:
Although one has the typical value $D^{\rm typ}_L \sim \exp(-\beta L^H)$,
we show via an exact analytical approach that the positive moments
($k>0$) scale like $\langle D^k_L
\rangle \sim \exp{[-c' (k \beta L^{H})^{1/(1+H)}]}$, and the negative
ones as $\langle D^{-k}_L \rangle \sim \exp(a' (k \beta L^{H})^2)$,
$c'$ and $a'$ being  numerical constants and $\beta$ the inverse
temperature. These results demonstrate that $D_L$ is strongly non-self-averaging.  
We further show that the
probability distribution of $D_L$ has a log-normal left tail and a
highly singular, one-sided log-stable right tail reminiscent of a Lifshitz singularity.
\end{abstract}
\maketitle
Transport in random media is extensively studied due to its practical and
fundamental importance \cite{bouch1,bouch2,chem}. In many cases, the
dynamics is modelled as a Langevin process, with a drift generated by a
quenched disordered potential. In theoretical analysis, the potential
landscape is taken to be either infinitely extended or periodic in space. 
Stochastic dynamics in a periodic potential, both random and deterministic, 
is commonly encountered in many different contexts, including 
modulated structures \cite{pol}, superionic conductors \cite{diet}, 
colloids in light fields \cite{evers,dean}, 
diffusion on regular \cite{rei,lac,kat,lind} 
and disordered \cite{katja,khoury,reim,khoury2,simon} solid surfaces, molecular motors on 
disordered tracks like  DNA/RNA \cite{kafri1,kafri2,kafri3},
and motion in a tilted potential due to a random polymer
\cite{salgado1}. 

Theoretical approaches often assume that the dynamics in a
periodic potential reproduces the behavior in an infinitely extended
potential. This is implemented by setting the period in the final result, e.g. for the velocity
(if any) or the diffusion coefficient, to infinity \cite{der,dsd}. It is crucial to investigate how far such an assumption holds. Especially in the context of numerical simulations carried out for
periodic systems, one may ask how reliably their results may be extrapolated to infinite systems.

\begin{figure}
\centering
\includegraphics[width=100mm]{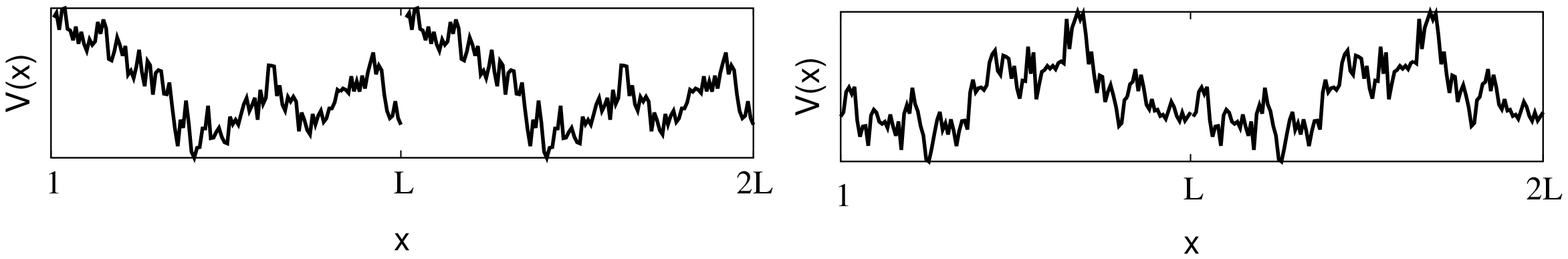}
\caption{Sketch of the potential $V(x)$ for the ratchet (left) and the
translationally invariant case (right). Here, $H=1/3$.}
\label{Fig0}
\end{figure}

In this work, we address these fundamental questions for a Langevin
dynamics $x(t)$ in a periodic, quenched random potential $V(x(t))$ [with $V(x(t) + L) = V(x(t))$]:
\begin{equation}
\eta \, \frac{d x}{dt} = - \frac{dV(x)}{dx} + \xi(t) \,,
\label{eom}
\end{equation} 
with $\eta$ the friction coefficient, $\xi(t)$ a Gaussian
white noise with zero mean and correlations
$\overline{\xi(t) \xi(t')} = 2 \eta T \delta(t-t')$, the overbar being
an average over the noise, and the temperature $T$ is in units of the
Boltzmann constant. We
consider two cases:\\
$\bullet$ the {\it ratchet} case where $V(x)$ is a fractional Brownian motion (fBm) in {\it time} $x \in
[0,L]$, with $V(0)=0$ and $V(L)$ arbitrary. Thus, $V(x)$ is a Gaussian process with zero mean, $\langle V(x)
\rangle=0$, and variance
\begin{equation}
\label{var}
\langle [V(x) - V(y)]^2 \rangle = \frac{V_0^2}{l^{2 H}} |x -
y|^{2 H} \,;\,\, x,y \in [0,L] \,,
\end{equation}
where $H \in (0,1)$ is the Hurst exponent, $V_0$ and $l$ define
respectively the typical amplitude of $V(x)$ and its scale of variation
over $x$. In (\ref{var}) the angular brackets denote averaging with respect to $V(x)$. 
Fig. \ref{Fig0} (left) shows a realization of $V(x)$, with a jump at $x=L$. \\
$\bullet$ the {\it translationally invariant} case, where $V(x)$ is a stationary
  Gaussian process, which at short length scales $|x-y| \ll L$ has the
  variance (\ref{var}), and satisfies $V(0)=V(L)$, so that all
  points are statistically equivalent. The particle in this case diffuses on a ring. A realization of such a $V(x)$ is
  shown in Fig. \ref{Fig0} (right).

The dynamics (\ref{eom}) involves a combination of two paradigmatic
situations: random motion in a periodic potential and the generalised Sinai
dynamics  in presence of
a force $F(x)  = - dV(x)/dx$ that is a time-independent stochastic variable with spatial correlations (except for $H=1/2$
when $V(x)$ is the trajectory of a Brownian
motion itself so that
(\ref{eom}) is the {\it periodic} Sinai model \cite{sinai}).
While the latter produces an archetypal subdiffusion with
logarithmically-confined trajectories, the periodicity of the random potential
enforces a long-time diffusive behavior with a diffusion coefficient
$D_L$. Here, we show 
that the trade-off between these two competing trends 
results in a rich statistical behavior of $D_L$. In particular, $D_L$ is strongly
 non-self-averaging, with both negative and positive moments exhibiting an anomalous dependence on the temperature, period $L$
and the order of the moment, and being supported by atypical
realizations of $V(x)$. For the ratchet case, we obtain exact analytical results, relying on exact bounds, for both positive and negative
moments of $D_L$. We also discuss the full form of the probability distribution of
$D_L$, and show that it is characterized by a log-normal left tail and a
highly singular log-stable right tail, reminiscent of a Lifshitz singularity. We finally
highlight the issue of sample-to-sample fluctuations of $D_L$. From standard scaling arguments and physical
intuition, one expects that our exact results for the ratchet case also hold for the translationally invariant situation, which is harder to analyze analytically. This is confirmed below by thorough numerical simulations~\cite{santachiara:2007}.

The dynamics (\ref{eom}) in an infinite system for arbitrary $H$, where $H>1/2$ ($H<1/2$) implies positively (negatively) correlated increments and superdiffusive (subdiffusive) $V(x)$, respectively, was discussed in
\cite{marinari} where it was shown that $\lim_{t \to \infty}\langle \overline{x^2(t)}
\rangle \sim \ln^{2/H}(t)$ (see also \cite{alberto}). In contrast, in a
periodic system, the long-time motion is diffusive for any given
realization of the potential $V(x)$, so that we have the diffusion
coefficient $D_L \equiv \lim_{t \to \infty}\overline{x^2(t)}/(2 t)$,
with $D_L$ given by 
\cite{c,a,leb,leb2,zw} (see also \cite{rei,dean,katja,baiesi})
\begin{equation}
\label{def}
\frac{D_L}{D^{(0)}}= \left(\int^L_0 \frac{dx}{L} \int^L_0 \frac{dy}{L}
\, e^{\beta [V(x) - V(y)]}\right)^{-1}\,,
\end{equation}
where $\beta$ is the inverse temperature, and $D^{(0)} = T/\eta$. Clearly, $D_L$ is a random
variable that fluctuates between realizations of $V(x)$, and has
support on $[0, D^{(0)}]$. The inverse of $D_L$ may be regarded as
a product of partition functions in potentials $V(x)$ and $-V(x)$, respectively. The Brownian
version of this quantity finds applications in disordered
systems and has been extensively studied, while our results for $H \ne 1/2$
apply to more general situations (note that the marginal case $H = 0$,
when $V(x)$ is log-correlated, was studied in \cite{Doussal}). The expression in Eq.~(\ref{def})
also describes the ground state energy in a toy model of
localization, and its average value was studied in Ref.
\cite{monthus} for $H
=1/2$.  

Turning to the discussion of the behavior of $D_L$, we first reduce the number of parameters.
In the following, we 
measure $L$ in units of $l$ [see Eq. (\ref{var})], absorb $V_0$ into $\beta$, and measure $D_L$
in units of $D^{(0)}$, so that $D_L$ has support
on $[0,1]$ \cite{foot_1}.  Now, 
the typical 
behavior of $D_L$ is easy to estimate as $D^{\rm typ}_L \propto L^2/\tau_{\rm typ}$,
where the dimensionless $L$ sets the scale of an inhomogeneous region,
and $\tau_{\rm typ}$ defines the typical (dimensionless) time a particle spends
in this region. The random potential $V(x)$ being a fBm, the typical
height of the potential barrier over a length $L$ scales like
$L^H$, for both the ratchet and the translationally invariant case. Assuming Arrhenius-type activation, one expects $\tau_{\rm typ}
\sim e^{\beta L^H}$, which implies 
\begin{eqnarray}\label{D_typ}
D^{\rm typ}_L \sim L^2 \exp{(- \beta L^H)} \;.
\end{eqnarray}
The average behavior of $D_L$ is a much more delicate question because, as 
can be seen from Eq.~(\ref{def}), computing the statistics of $D_L$ is a highly non-trivial task 
that involves the study of an exponential functional of fBm, for which standard methods like the 
Feynman-Kac formula are of little use for $H \neq 1/2$. 
\begin{figure}[!h]
\centering
\begin{tabular}{lr}
\parbox[l]{4.5cm}{
\includegraphics[width=43mm]{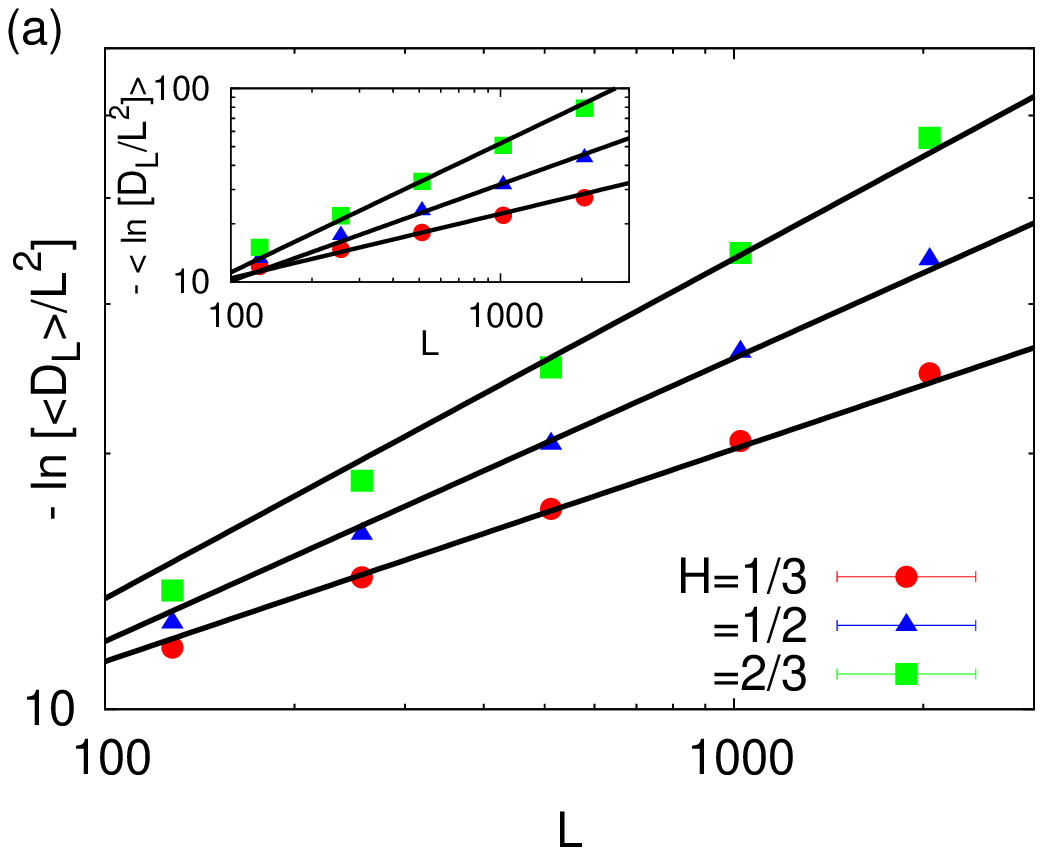}
}&
\parbox[r]{3.5cm}{
\includegraphics[width=37mm]{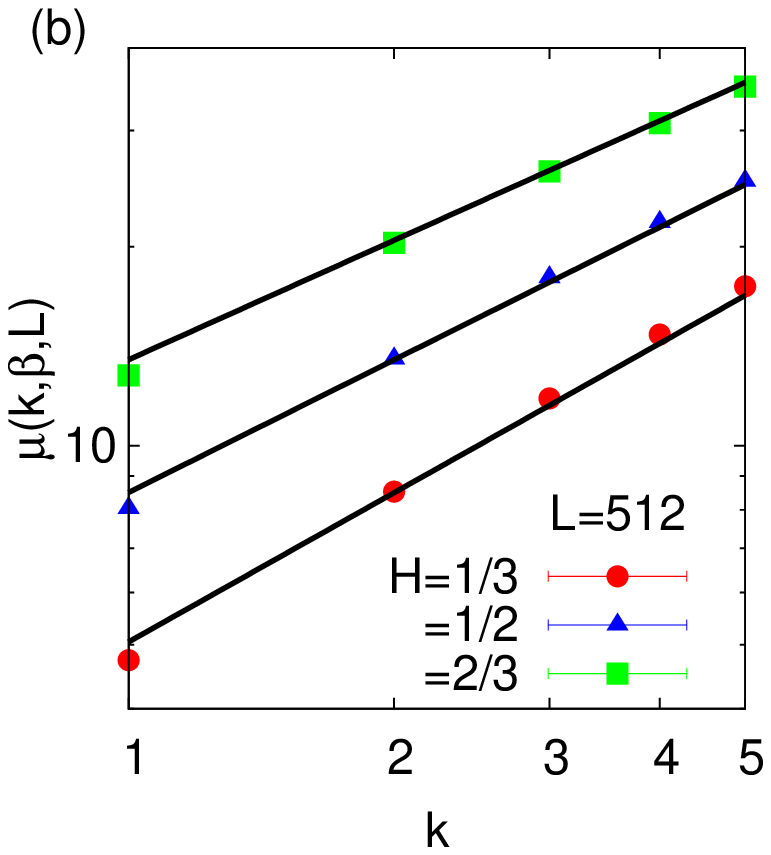}
}
\end{tabular}
\caption{(Color online) (a) $- \ln[\langle D_L\rangle/L^{2}]$ versus $L$ for three values of $H$ corresponding to
diffusive, subdiffusive, and superdiffusive $V(x)$.
The inset shows $-\langle \ln [D_L/L^2] \rangle$ as a function of $L$
for three values of $H$, see Eq.~(\ref{log}). (b) $\mu(k,\beta,L)$ in Eq.~(\ref{mu}) as a function
of $k$ for $L=512$ and  
three values of $H$. In all cases, the symbols denote simulation results
for the translationally invariant $V(x)$, while the slopes of the solid
lines correspond to the results derived for the ratchet case.}
\label{Fig1}
\end{figure}

Let us first summarize our main analytical results obtained for the ratchet case: We find that the average of the
logarithm of $D_L$ is given, to
leading order in $L$, by
\begin{equation}
\label{log}
\Big\langle \ln D_L\Big\rangle  \propto - 2 m \beta L^H \,,
\end{equation}
with $m = \langle {\rm max}_{s \in [0,1]} V(s)\rangle$.  The result in Eq.~(\ref{log}) is consistent with the 
 logarithmic growth of the disorder-averaged mean-square
 displacement in an infinite system: $\lim_{t \to \infty}\langle \overline{x^2(t)}\rangle
 \sim \ln^{2/H}(t)$, and the argument leading to Eq.~(\ref{D_typ}).
Further on, we obtain sharp bounds for the positive moments $(k > 0)$ of
the random variable $D_L$:
\begin{equation}
\label{bounds}
A_k(L) \leq \Big\langle D^k_L\Big\rangle \leq B_k(L) \,,
\end{equation} 
where, in the limit $L \to \infty$, the bounds satisfy 
\begin{equation}
A_k(L) = \exp\left[- (1+H) \left(2k \beta\right)^{\frac{1}{1+H}}
\left(\frac{C}{H} L\right)^\frac{H}{1+H}\right], B_k(L) =  \exp\left[- (1+H) \left(k \beta \right)^{\frac{1}{1+H}}
\left(\frac{c}{H} L\right)^\frac{H}{1+H}\right] \,,
\end{equation}
with constants $c$ and $C$, $0 < c \leq C < \infty$, being independent
of $L$, $k$ and
$\beta$. Both bounds exhibit the same dependence on $L,k,\beta$, from which we infer that the exact asymptotic result has the same
functional form. 
Finally, for the negative moments of $D_L$, we find 
\begin{equation}
\label{neg}
\Big\langle D^{-k}_L\Big\rangle \sim \exp\left( \frac{a ^2 k^2 \beta^2 L^{2 H}}{4}\right) \,,
\end{equation}
where $a$ is a constant (independent of $L, k$ and $\beta$). 
Our exact results,
Eqs.~(\ref{bounds}) and~(\ref{neg}), thus show that the positive and negative moments are
dominated by atypical realizations of $V(x)$, in contrast to $\langle
\ln D_L
\rangle$, see Eq.~(\ref{log}). 

We now turn to a derivation of our results. Using Eq.~(\ref{def}), the
logarithm of $D_L$ can be formally written as
$\ln D_L =  \ln J_+(L) +  \ln J_-(L) + 2 \ln L$,
where
 $J_\pm(L)$ are stationary currents through a finite sample
 of length $L$ with potentials $\pm V(x)$
 \cite{alberto}: $J_{\pm}(L) = \Big[\int^L_0 dx \, \exp[\pm
 \beta V(x)]\Big]^{-1}$. 
Statistical properties of these currents for the Sinai problem ($H = 1/2$) are known \cite{gleb,gleb2,cecile,alain,sid,schehr}. 
Using the results of \cite{molchan}, and noting that with $\langle V(x)
\rangle=0$, $J_{+}(L)$ and $J_{-}(L)$ have equal moments,
we have for arbitrary $H$ and to leading order in $L$, $
\Big\langle \ln J_+(L)\Big\rangle = \Big\langle \ln J_-(L)\Big\rangle
\propto - m \beta L^H$, which yields Eq.~(\ref{log}).

The proof of the result in Eq.~(\ref{bounds}) is based on a Theorem due to Monrad and Rootz\'en \cite{monrad} on the
probability that a fBm $V(x)$, with $V(0)=0$, remains within a strip of width $\epsilon$ for the time $x \in
[0,L]$. Defining $M_L \equiv \max_{0 \geq x \geq L}|V(x)|$,
the Monrad-Rootz\'en Theorem, in our notation, states that
$P(M_L \leq \epsilon)$ satisfies 
\begin{equation}
\label{monrad}
\exp\left( - C  {L}\,{\epsilon^{-\frac{1}{H}}} \right) \leq P(M_L \leq \epsilon) \leq \exp\left( - c {L}\,{\epsilon^{-\frac{1}{H}}} \right),
\end{equation}
for $0 < \epsilon \leq L^H$, $c$ and $C$ being $L$-independent constants
[see Eq.~(\ref{bounds})].  

Consider the lower bound in Eq.~(\ref{bounds}). Suppose we average the
positive definite quantity $D^k_L$ by considering instead of the entire
set $\Omega$ of all possible paths $V(x)$ only a  subset $\Omega' \subset \Omega$ of
paths such that $M_L \leq \epsilon$.
This gives the lower bound $
\Big\langle D^k_L \Big\rangle_{\Omega} \geq \Big\langle
D^k_L\Big\rangle_{\Omega'} P(M_L\le \epsilon)$. However, for paths in $\Omega'$, we have
$\exp(\beta[V(x) - V(y)]) \leq \exp(2\beta \epsilon)$, and hence,
$
\left(\int^L_0 \frac{dx}{L} \int^L_0 \frac{d y}{L} e^{\beta [V(x) -
V(y)]}\right)^{-1} \geq e^{- 2\beta \epsilon}$.
Therefore, we obtain $
\Big\langle D^k_L \Big\rangle_{\Omega} \geq  e^{- 2k \beta \epsilon} \,
P(M_L \leq \epsilon)$.
Making the inequality more stringent by choosing the lower bound in Eq.~(\ref{monrad}), we get
\begin{equation}
\label{pl}
\Big\langle D^k_L\Big\rangle_{\Omega} \geq e^{- 2k \beta \epsilon} \,  \exp\left( - C {L}{\epsilon^{-\frac{1}{H}}} \right) \,,
\end{equation}
which holds for any $\epsilon$ with $0 < \epsilon \leq L^H$.

The function on the right hand side (rhs) of Eq.~(\ref{pl}) is a
non-monotonic function of $\epsilon$, attaining its maximum 
at $\epsilon = \epsilon_{\rm opt}= (C L/k 2\beta
H)^{H/(1+H)}$. Clearly, the best lower bound corresponds to the choice
$\epsilon = \epsilon_{\rm opt}$, leading to the lower bound in Eq.~(\ref{bounds}).
Note that to satisfy the conditions of validity of the Monrad-Rootz\'en
Theorem, we require that $\epsilon_{\rm opt} \leq L^H$, that is, $2k \beta
L \geq C/H$, which is easily realized for sufficiently large $L$.
The derivation of the lower bound is an example of the  Lifshitz {\it optimal fluctuation} method \cite{lifshitz}, which has been used to bound the survival probability of particles diffusing in the presence of randomly scattered immobile traps (see, e.g., \cite{gras}).

We now discuss the derivation of the upper bound. To this end, we
discretize $x$, and write the rhs of Eq.~(\ref{def}) as
\begin{equation}
\int^L_0 \frac{dx}{L} \int^L_0 \frac{d y}{L} e^{\beta [V(x) - V(y)]}
\sim \sum_{i,j=1}^N e^{\beta [V(j) - V(i)]} \,.
\end{equation}
Given that the fBm starts at $V(0)=0$, at least one term in the double
sum on the rhs takes the value
$\exp(\beta M_L)$, corresponding to the point $x=0$ and the point where $|V(x)|$ attains 
its maximal value $M_L$. Now, as all the other terms are positive, we have the bound
$D^{-1}_L \geq \exp(\beta M_L)$, and, thus,
\begin{equation}
\langle D^k_L \rangle \leq \Big\langle e^{- k \beta
M_L}\Big\rangle= k\beta \int^{\infty}_0 d\epsilon P(M_L \leq \epsilon)
e^{- k \beta \epsilon}\,.
\label{upper-bound}
\end{equation}
The integral in the rhs is dominated, for large $L$, by the small $\epsilon$ region, where we can thus use the upper bound in (\ref{monrad}).
Performing the remaining integral over $\epsilon$ by the saddle-point method and omitting the pre-exponential terms lead to the upper bound in Eq.~(\ref{bounds}). The result in (\ref{bounds}) has several striking features. Namely, the function
\begin{equation}
\label{mu}
\mu(k,\beta, L) = - \ln\Big\langle D^k_L\Big\rangle \,,
\end{equation}
(as compared to its typical counterpart given by Eq.~(\ref{D_typ}) as $\mu_{\rm typ}(k,\beta,L)  \sim k \beta L^H$),
(a) grows sub-linearly with $k$ (multifractality), (b) is a non-analytic, sublinear function of $\beta$, which implies a rather unusual 
sub-Arrhenius dependence of the positive moments on the temperature, and
(c) exhibits a slower anomalous growth with $L$ as $\sim L^{\frac{H}{1
+ H}}$. This means that the disorder-averaged $D_L$ is generically larger than the one expected on the basis of
typical realizations of the disorder. 
In turn, this implies that the behavior of the positive moments of 
$D_L$ is supported by atypical realizations of
disorder, reminiscent of the so-called Lifshitz singularities, as
discussed above. In conclusion, one cannot infer the dynamical behavior
in an infinite system from the positive moments of $D_L$. This is surprising
at first glance, as $\langle\overline{x^2(t)}\rangle$ is linearly
proportional to $\langle D_L\rangle$, and shows that 
the limits $t \to \infty$ and $L \to \infty$ do
not commute in this system.  

\begin{figure}[!t]
\centering
\includegraphics[width=70mm]{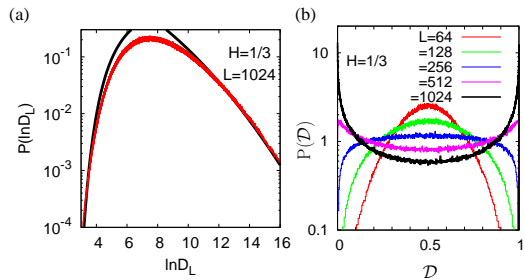}
\caption{(Color online) (a) Distribution $P(\ln D_L)$ for $L=1024$ and
$H=1/3$. The red line denotes numerical results, while analytical predictions for the
right and left tails behaving as $\exp(-a(H)/\ln^{1/H}D_L^{-1})$ and
$\exp(-b(H) \ln^2(D_L))$, respectively, with $a(H)$ and $b(H)$ being constants, are shown by black
lines. (b) Numerical results for $P({\cal
D})$ for different $L$ and $H=1/3$.}
\label{Fig2}
\end{figure}

The behavior of the negative moments $\langle D^{-k}_L\rangle$ with  $k = 1,2, \dots$
is determined by essentially the same approach as above. Note that both the lower and the upper bound on
$D^{-k}_L$ are made tighter for a given realization of $V(x)$ by using
$D^{-k}_L \sim
\exp(k\beta S)$, where $S$ is the span of $V(x)$ (the difference between the maximum and minimum) on the interval $[0,L]$. Therefore, in contrast to the positive moments, the negative moments are supported by realizations of $V(x)$ with a large span. Using the result
that for large $S$, $P(M_L = S) \sim \exp(- S^2/a L^{2 H})$,
with $a$ a constant, integration of
Eq.~(\ref{upper-bound}) gives the result announced in Eq.~(\ref{neg}),
which displays a super-Arrhenius dependence on the temperature, a
superlinear dependence on $k$, and a strong dependence on $L$.
A similar result was obtained earlier in \cite{1}. 
We note that, as
for the positive moments, one cannot deduce the behavior in an
infinite system from that of negative moments of $D_L$ in a periodic
system, as the latter is supported by atypical realizations of $V(x)$ that have anomalously large span scaling as $S \sim L^{2 H}$, while the
typical behavior is $S_{\rm typ} \sim L^H$. 

Based on our results for the moments, we now obtain the
probability distribution $P(D_L)$. As already explained, the
behavior of the negative moments is supported by anomalously stretched
trajectories of $V(x)$ for which the value of $D_L$ is small. One may
thus expect in view of the form of the moments in Eq.~(\ref{neg}) that
for small $D_L$, $P(D_L)$ is log-normal:
\begin{equation}
\label{ln}
P(D_L) \sim \frac{1}{a \beta L^H D_L} \exp\left(- \frac{\ln^2(D_L)}{a^2 \beta^2 L^{2 H}}\right) \,.
\end{equation}  
To analyze the behavior of $P(D_L)$ for $D_L$ close to $1$, we recall
the formal definition of the one-sided L\'evy distribution
${\cal L}_{\nu}(z)$, $0 \leq z < \infty$, of order $\nu$ (see, e.g.,
\cite{led}):
\begin{equation}
\label{levy}
\int^{\infty}_0 dz e^{- p z} {\cal L}_{\nu}(z) = e^{-p^{\nu}} \,.
\end{equation}
The asymptotic behavior of ${\cal L}_{\nu}(z)$ is well-known \cite{led}, and, in particular, one has ${\cal L}_{\nu}(z) \sim
z^{-\sigma} \exp(-b/z^{\tau})$ for $z \to 0$, where $b$ is a computable constant,
$\sigma = (2 - \nu)/(2 (1-\nu))$, and $\tau = \nu/(1-\nu)$. It is
important to note that this precise asymptotic form is responsible for
the stretched-exponential behavior in Eq.~(\ref{levy}), which is
immediately verified by substituting the form in Eq.~(\ref{levy}), and
performing the integration by the saddle-point method. Moreover, one
realizes by making in Eq.~(\ref{levy}) a change of the integration
variable $z = \ln(1/D_L)/\beta L^H$, choosing $\nu = 1/(1+H)$, and
setting $p = k/\beta L^H$ that Eq.~(\ref{levy}) becomes identical to
the result in Eq.~(\ref{bounds}), up to numerical factors.
It follows that for $D_L$ close to $1$ (i.e., $z$ close to $0$), the distribution function behaves as
\begin{equation}
\label{le}
P(D_L) \sim \frac{1}{\beta L^H D_L} {\cal
L}_{\frac{1}{1+H}}\left[{\ln(D_L^{-1})}/{\beta L^H}\right] \,.
\end{equation}
Using the asymptotic ${\cal L}_\nu$ given above, we get
that $P(D_L)$ is highly singular near the right edge, $P(D_L) \sim \exp[-b \beta^{1/H}
L/(1-D_L)^{1/H}]$, similar to the Lifshitz singularity. 

We now consider the translationally invariant case. Here, we expect our above analysis, in
particular, result (\ref{bounds}) to hold, up to possible numerical factors. To
demonstrate this, we now present results of extensive numerical
simulations: Fig. \ref{Fig1} for $- \ln(\langle
D_L\rangle/L^{2})$, $-\langle \ln (D_L/L^2) \rangle$, and
$\mu(k,\beta,L)$, and Fig. \ref{Fig2}(a) for $P(\ln D_L)$ indeed show a very good agreement
that supports our expectations. 

To close, we ask: if we have two different
realizations of $V(x)$, and correspondingly, two different values,
$D_L$ and $D'_L$, of the diffusion coefficient, how likely are
these values equal? We introduce a random variable  ${\cal D}\equiv
\frac{D_L}{D_L + D'_L} \,, \,\, {\cal D}
\in [0,1]$,
and analyze its distribution $P({\cal D})$ via numerical simulations.
 Clearly, ${\cal D} = 1/2$ maximizing $P({\cal D})$ implies that the two
 values of $D_L$ are most likely very close to one another. 
 Variables such as ${\cal D}$ play a key role in various
  scale-independent hypothesis testing procedures, in
  classical problems in statistics, in
  signal processing (see, e.g., \cite{mac}), and 
  in the analysis of chaotic scattering in few-channel systems
  \cite{greg}. Such variables are used to characterize 
the effective width of {\it narrow} distributions possessing moments of arbitrary order \cite{carlos,thiago}.      

In Fig. \ref{Fig2}(b), we present numerical results for $P({\cal D})$ for
 different values of $L$ and $H=1/3$, for the translationally invariant case. We observe an interesting phenomenon of
  a change in the form of the distribution as $L$ is increased. For
  relatively small $L$, the distribution is bell-shaped and centered at
  ${\cal D}=1/2$. However, on increasing $L$,
  $P({\cal D})$ broadens, becomes almost flat at a certain critical $L$
  (whose value depends on $H$), and then changes its shape so that
  ${\cal D} = 1/2$ minimizes the distribution. This implies that for
  sufficiently large $L$, two values of the diffusion coefficients
  obtained for two different realizations of $V(x)$ are most likely very
  different, and the event $D_L = D'_L$ is the least probable. Note that a similar
  dependence in the distribution of the prefactor
  in the Sinai law on the strength of disorder was recently observed in~\cite{boyer}.

GS is supported by the ANR
grant 2011-BS04-013-01 WALKMAT. This project is partially supported by the Indo-French Centre for the Promotion of
Advanced Research under Project 4604-3. SG, AR and GS thank the Galileo
Galilei Institute for Theoretical Physics, Florence, Italy for the
hospitality and the INFN for partial support during the completion of
this work.


\begin{thebibliography}{99}
\bibitem{foot_1} $l, V_0$ and $D^{(0)}$ may be made explicit in the final result by the replacements $L \to L/l$, $\beta \to \beta V_0$ and
$D \to D/D^{(0)}$.
\end{thebibliography}
\end{document}